\documentclass[aip,apl,twocolumn,12pt,reprint]{revtex4-1}

\draft 
\usepackage{graphicx}
\usepackage{amsmath}
\begin{document}


\title{Enhanced Inverse Spin-Hall Effect in Ultrathin Ferromagnetic/Normal Metal Bilayers} 



\author{T.D. Skinner}
\affiliation{Cavendish Laboratory, University of Cambridge, CB3 0HE, United Kingdom}

\author{H. Kurebayashi}
\affiliation{Cavendish Laboratory, University of Cambridge, CB3 0HE, United Kingdom}
\affiliation{PRESTO, Japan Science and Technology Agency, Kawaguchi 332-0012, Japan}

\author{D. Fang}
\affiliation{Hitachi Cambridge Laboratory, Cambridge, CB3 0HE, United Kingdom}

\author{D. Heiss}
\altaffiliation[Current address: ]{Department of Electrical Engineering, Technische Universiteit Eindhoven, 5600 MB Eindhoven, The Netherlands}
\affiliation{Cavendish Laboratory, University of Cambridge, CB3 0HE, United Kingdom}

\author{A.C. Irvine}
\affiliation{Cavendish Laboratory, University of Cambridge, CB3 0HE, United Kingdom}

\author{A.T. Hindmarch}
\altaffiliation[Current address: ]{Centre for Materials Physics, Durham University, DH1 3LE, United Kingdom}
\affiliation{School of Physics and Astronomy, University of Nottingham, NG7 2RD, United Kingdom}

\author{M. Wang}
\affiliation{School of Physics and Astronomy, University of Nottingham, NG7 2RD, United Kingdom}

\author{A.W. Rushforth}
\affiliation{School of Physics and Astronomy, University of Nottingham, NG7 2RD, United Kingdom}

\author{A.J.Ferguson}
\email[Electronic mail: ]{ajf1006@cam.ac.uk}
\affiliation{Cavendish Laboratory, University of Cambridge, CB3 0HE, United Kingdom}

\date{\today}

\begin{abstract}
We measure electrically detected ferromagnetic resonance in microdevices patterned from ultra-thin Co/Pt bilayers. Spin pumping and rectification voltages are observed and distinguished via their angular dependence. The spin-pumping voltage shows an unexpected increase as the cobalt thickness is reduced below 2 nm. This enhancement allows more efficient conversion of spin to charge current and motivates a theory modelling the dependence of impurity scattering on surface roughness.
\end{abstract}


\maketitle 


Ferromagnetic/heavy metal (e.g. Co/Pt) bilayers provide a model system in which to study spin transfer phenomena. A charge current in the heavy metal causes diffusion of a spin current into the ferromagnet via the spin hall effect \cite{Kato10122004, PhysRevLett.94.047204}. The resulting angular momentum transfer can either change the amplitude of magnetisation precession induced by conventional ferromagnetic resonance\cite{PhysRevLett.101.036601} or directly drive magnetisation precession\cite{PhysRevLett.106.036601}. In addition, switching of a nanoscale magnetic element has been achieved, indicating that the spin-Hall effect may be used to control memory elements\cite{Liu04052012}. Conversely, the precessing magnetisation in the ferromagnetic layer drives a spin current into the heavy metal layer\cite{PhysRevLett.88.117601,PhysRevB.66.224403}, where the inverse spin-Hall effect\cite{NATURE20064427099,saitoh:182509} converts it into a measureable charge current. This process, known as spin-pumping, has become a common laboratory technique to create spin currents in diverse materials
\cite{PhysRevLett.97.216603,Nature:Saitoh,PhysRevLett.104.046601,PhysRevLett.107.046601,PhysRevLett.107.066604,Nature1092010}.
A charge current in ultra-thin Co/Pt bilayers has also been reported to act on the magnetisation via a `Rashba' spin-orbit torque\cite{2010NatMa.230M, 2011NatMa.419M}, due to a relativistic magnetic field existing at the heavy metal interface.
In this Letter, in contrast to previous research on thicker layers\cite{PhysRevB.83.144402,PhysRevB.85.144408}, we investigate spin-pumping in ultra-thin Co/Pt bilayers in which the interface region is a significant proportion of the bulk ferromagnet and Pt layers. By keeping the platinum layer thickness constant, we eliminate any variation in the bulk inverse spin-Hall detection. We examine the strength of the spin-pumping voltage in the platinum layer as we vary the thickness of the ferromagnet.

In our study the samples are thin bars of Co/Pt with nominal cobalt thickness $d_{\mathrm{Co}}=$ 1, 1.25, 1.5, 1.75 and 2 nm capped with a 3 nm Pt layer. From x-ray reflectivity (XRR) measurements we estimate the uncertainty in the thickness of these layers to be 10\%. An out-of-plane microwave magnetic field ($h_{z}e^{i\omega t}$), for ferromagnetic resonance (FMR), was generated over the sample area by an on-chip coplanar stripline, shorted 1 $\mu$m away from the sample.  
\begin{figure}
 \includegraphics{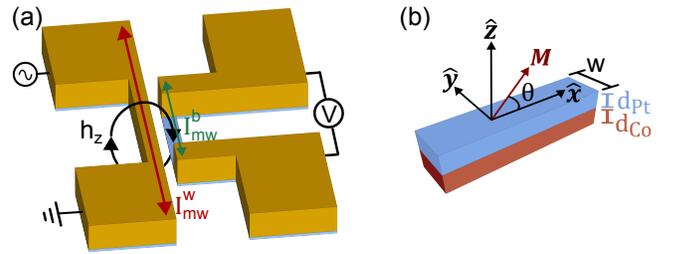}
 \caption{\label{fig:device}(Colour online) (a) Measurement schematic showing coplanar stripline on left with microwave current, $I_{\mathrm{mw}}^{\mathrm{w}}$, generating a perpendicular microwave field over the bar area. A microwave current, $I_{\mathrm{mw}}^{\mathrm{b}}$, is coupled into the bar. The voltage is measured across the bar contacts with a lock-in amplifier.
 (b) The bar consists of a Pt layer deposited on top of a cobalt layer. The in-plane angle $\theta$ is defined as the angle between the bar direction and the magnetisation.}
\end{figure}
The devices were fabricated from films sputtered on thermally oxidised silicon. Electron beam lithography was used for patterning, and then 1 x 10 $\mu$m bars and adjacent striplines were defined with Ar ion-milling. The bars are contacted by 200 nm thick gold contacts which were deposited by evaporation at the same time as the gold striplines. A schematic of the device and measurement is shown in figure \ref{fig:device}.

The sample was mounted on a low loss printed circuit board (PCB). A 15 GHz microwave signal was sent via a coaxial cable into the PCB waveguide and then into the shorted stripline to ground. The signal power in the coaxial cable directly before the PCB was 14.5 dBm. As the PCB waveguides and on-chip striplines are identical for each device, we expect similar microwave currents, $I_{\mathrm{mw}}^{\mathrm{w}}$, in every stripline. In this measurement we assume the microwave field generated is identical for each sample.

The microwave signal was pulse modulated at low frequency (23.45 Hz) allowing a lock-in amplifier to detect the DC voltage ($V_{\mathrm{dc}}$) across the sample.  The sample was positioned in a 3-axis vector magnet at a temperature of 250K. For a particular direction, the external magnetic field was swept from high to low field, and the ferromagnetic resonance was observed as a combination of symmetric and antisymmetric Lorentzian peaks in $V_{\mathrm{dc}}$.

$V_{\mathrm{dc}}$ is thought to be generated through two effects: the inverse-spin-Hall effect (ISHE) and rectification. 
During steady-state precession, the driving torque is balanced by a damping torque. The Pt layer adjacent to the ferromagnet is an efficient spin-current sink and contributes to the damping by transferring angular momentum between the Co and Pt layers via a spin-current. The spin-current, $j_{\mathrm{s}}$, injected into the Pt layer through the ISHE generates a transverse charge current given by\cite{PhysRevB.85.144408}
\begin{equation}\label{eq:jc}
\mathbf{j_{\mathrm{c}}}=\theta_{\mathrm{ISHE}}\left(\frac{2e}{\hbar}\right)\mathbf{j_{\mathrm{s}}}\times\boldsymbol{\sigma}
\end{equation}
An initial spin current $j_{\mathrm{s}}^{\mathrm{0}}$ at the interface decays due to spin relaxation as it penetrates the Pt layer. This yields a total charge current of $I_{\mathrm{c}}$ which creates a voltage $V_{\mathrm{ISHE}}=I_{\mathrm{c}}R$ across the bar. $\theta_{\mathrm{ISHE}}$, $e$, $\hbar$ and $\sigma$ represent the spin-Hall angle, the elementary charge, the reduced Planck constant and the spin-polarisation vector of the spin-current respectively. 

The microwave current in the shorted stripline can couple into the sample, to give another microwave current, $I_{\mathrm{mw}}^{\mathrm{b}}$. At resonance the magnetisation will precess at the same frequency as this current. Precession of the magnetisation causes an oscillating component to the resistance, due to the anisotropic magnetoresistance (AMR) $R=R_{0}+\Delta R\cos^{2}\theta$. This multiplies with the microwave current to give a measurable $V_{\mathrm{dc}}$. Combining this with  $V_{\mathrm{ISHE}}$, the real part of the voltage is given by the sum of symmetric and antisymmetric parts\cite{PhysRevB.78.104401,PhysRevB.85.214423,PhysRevB.85.144408}
\begin{align}\label{V_dc}
&V_{\mathrm{dc}}&=&\left(V_{\mathrm{AMR}}\cos\phi + V_{\mathrm{ISHE}}\right)\frac{\Delta H^{2}}{\left(H-H_{0}\right)+\Delta H^{2}} \nonumber\\&
&+&V_{\mathrm{AMR}}\sin\phi\frac{\Delta H\left(H-H_{0}\right)}{\left(H-H_{0}\right)+\Delta H^{2}}
\end{align}
with $V_{\mathrm{AMR}}$ and $V_{\mathrm{ISHE}}$ given by
\begin{equation}
V_{\mathrm{AMR}}=\frac{1}{2}I_{\mathrm{mw}}^{\mathrm{b}}\Delta RA_{xx}\sin\left(2\theta\right)h_{z}
\end{equation}
\begin{equation}\label{eq:V_ISHE}
V_{\mathrm{ISHE}}=I_{\mathrm{c}}R=\theta_{\mathrm{SH}}w\left(\frac{2e}{\hbar}\right)\lambda_{\mathrm{sd}}\tanh\left(\frac{d_{\mathrm{Pt}}}{\lambda_{\mathrm{sd}}}\right)j_{\mathrm{s}}^{\mathrm{0}}R\sin\theta
\end{equation}
In these expressions, $H$ is the externally applied magnetic field, $H_{\mathrm{0}}$ is the resonant field and $\Delta H$ is the linewidth of the resonance. $\phi$ is the phase difference between the coupled current and the magnetisation precession. $d_{\mathrm{Pt}}$ and $w$ are the thickness of the platinum layer and the width of the bar. $\Delta R$, $R$ and $\lambda_{\mathrm{sd}}$ are the AMR coefficient, the sample resistance and the spin diffusion length in Pt respectively. $A_{xx}$ is related to the diagonal term of the AC magnetic susceptibility by $\chi_{xx}/M_{\mathrm{S}}$, where $M_{\mathrm{S}}$ is the saturation magnetisation\cite{PhysRevB.76.224430}. The magnetisation always lies in the plane of the sample due to the demagnetisation field and the negligible in-plane magnetic anisotropy.

 \begin{figure}
 \includegraphics[]{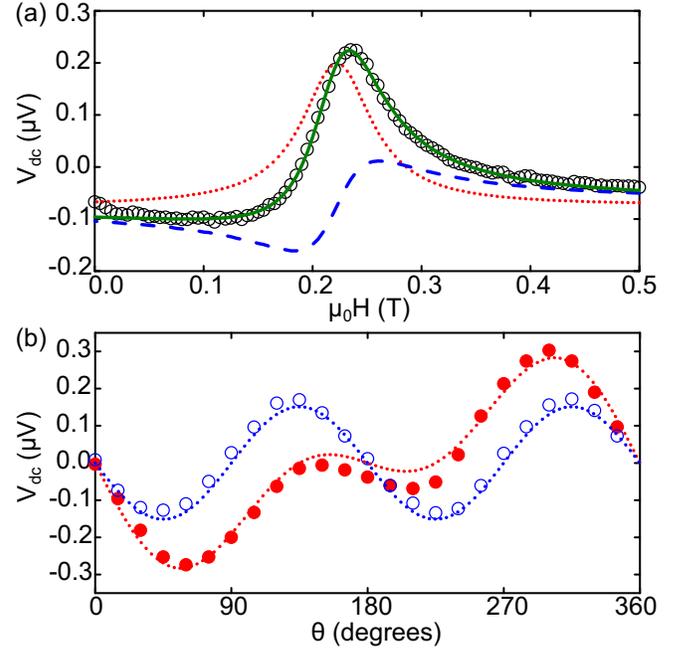}
 \caption{\label{fig:lorentz}(Colour online) (a) Detected voltage for a 2 nm device for a single field sweep. The FMR peak is fitted (solid green line) by a combination of symmetric (dotted red line) and antisymmetric (dashed blue line) Lorentzian curves.
 (b) The angular dependences of the symmetric (full red circles) and antisymmetric (open blue circles) voltages are each fitted by a linear combination of $\sin\theta$ and $\sin 2\theta$ terms.}
 \end{figure}

Only rectification can produce an antisymmetric Lorentzian, as the phase information needed to produce the asymmetry is held in the relative phase of the resistance and microwave current. Also observe that the two detection mechanisms have different angular dependencies, which allows us to distinguish them. The rectification voltage is proportional to $\sin 2\theta$ due to the symmetry of the AMR, whereas the angular dependence of the ISHE, given by the cross product in equation \ref{eq:jc}, makes the spin-pumping signal proportional to $\sin\theta$.

We measured FMR resonances for a series of in-plane angles and fitted the symmetric and antisymmetric Lorentzian peaks (see figure \ref{fig:lorentz}a), defining $V_{\mathrm{sym}}$ and $V_{\mathrm{asy}}$ as the coefficients of the symmetric and antisymmetric peaks in equation \ref{V_dc}. The angular dependencies of both the symmetric and antisymmetric terms are fitted well by a combination of $\sin\theta$ and $\sin 2\theta$ components. Figure \ref{fig:lorentz}b shows the fitting for a sample with a 2 nm Co layer. Neither of the detection methods proposed explains the antisymmetric $\sin\theta$ component. This component is only significant in the 1 nm Co layer.
 
 \begin{figure}
 \includegraphics[]{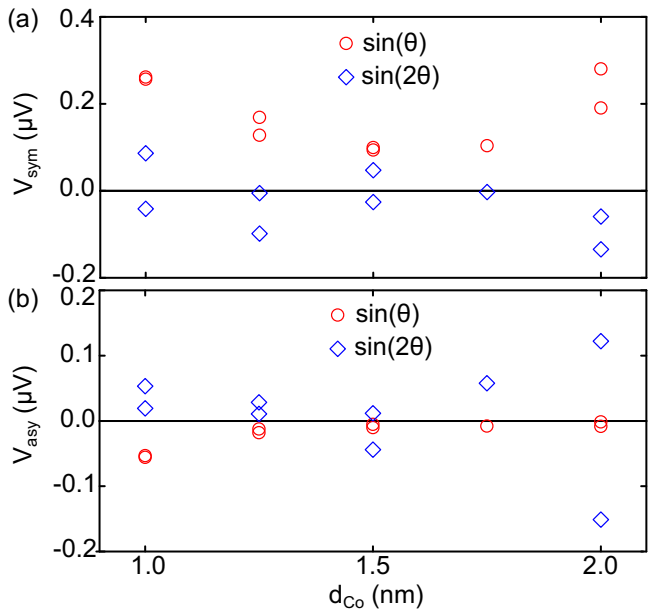}
 \caption{\label{fig:thickness}(Colour online) (a),(b) Cobalt thickness dependence of the fitted symmetric and antisymmetric $\sin\theta$ (blue diamonds) and $\sin 2\theta$ (red circles) voltage components.}
 \end{figure}

We repeated the measurements for the five cobalt thicknesses, using identical device structures, and the same experimental parameters. We also repeated measurements in a second device for all cobalt layer thicknesses except 1.75 nm to show the variation between devices. Figure \ref{fig:thickness} shows the detected voltages against cobalt thickness. Whilst there is a clear trend in the $\sin\theta$ components of both voltage parts, the $\sin 2\theta$ components are not consistent in magnitude or sign even between devices from the same layer structure. We attribute this to variation in the relative phase of the microwave current coupled into each device bilayer, $I_{\mathrm{mw}}^{\mathrm{b}}$, and the microwave current in the coplanar stripline generating the magnetic field, $I_{\mathrm{mw}}^{\mathrm{w}}$. As the device and coplanar stripline microstructures are nearly identical, we expect the amplitude and phase of $I_{\mathrm{mw}}^{\mathrm{b}}$ to be dominated by the milli-scale arrangement of bond wires and pads, which do vary between devices. The bond-wire lengths ($\sim$2 mm) are close to the free-space wavelength (20 mm) and could act as an antenna, coupling microwave current into the device bilayer. Unlike the rectification signal, the spin-pumping signal is insensitive to $I_{\mathrm{mw}}^{\mathrm{b}}$ and consequently is reproducible between devices.

The spin-current injected into the platinum layer is dependent on both the Gilbert damping and effective magnetisation which are themselves dependent on the cobalt thickness:\cite{PhysRevB.85.144408}
\begin{equation}\label{eq:js}
j_{\mathrm{s}}^{0}=\frac{g_{\mathrm{eff}}^{\uparrow\downarrow}\gamma h_{z}^{2}\hbar}{8\pi}\frac{\mu_{\mathrm{0}}M_{\mathrm{eff}} +\sqrt{\left(\mu_{\mathrm{0}}M_{\mathrm{eff}}\right)^2 +4\frac{\omega^2}{\gamma^2}}}{\alpha_{\mathrm{eff}}^2\left(\left(\mu_{\mathrm{0}}M_{\mathrm{eff}}\right)^2+4\frac{\omega^2}{\gamma^2}\right)}
\end{equation}
Here, $g_{\mathrm{eff}}^{\uparrow\downarrow}$ is the spin-mixing conductance, $\gamma$ is the gyromagnetic ratio and $M_{\mathrm{eff}}$ is the effective magnetisation.
The effective Gilbert damping constant, $\alpha_{\mathrm{eff}}$, has a contribution, not only from the volume of the ferromagnet, but also from the spin pumping at the interface\cite{PhysRevB.85.144408}
\begin{equation}\label{eq:alpha}
\alpha_{\mathrm{eff}}=\alpha_0 + \frac{g\mu_{\mathrm{B}}g_{\mathrm{eff}}^{\uparrow\downarrow}}{M_{\mathrm{S}}d_{\mathrm{Co}}}
\end{equation} 

Likewise, the effective magnetisation has a bulk contribution from the demagnetisation field, but also from a perpendicular uniaxial anisotropy originating from the interface\cite{Draaisma1987351}
\begin{equation}\label{eq:Meff}
M_{\mathrm{eff}}=M_{\mathrm{S}} - \frac{H_{\mathrm{U}}^{\mathrm{int}}}{d_{\mathrm{Co}}}
\end{equation}

\begin{figure}[]
 \includegraphics[]{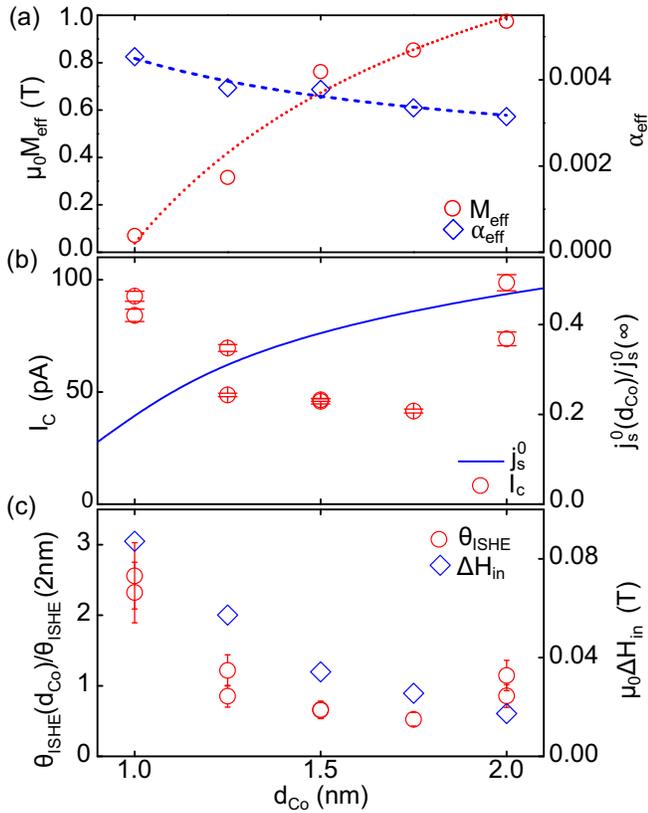}
 \caption{\label{fig:current}(Colour online) (a) By measuring FMR out-of-plane of the sample and self-consistently fitting the magnetisation angle and resonant field to the Kittel and energy equations, we determine the effective magnetisation in each sample \cite{ando:103913}. Measured values of $M_{\mathrm{eff}}$ (red circles) are fitted well by equation \ref{eq:Meff} (dotted line).
We calculate the Gilbert damping by measuring the frequency dependence of the linewidth, $\Delta H = \Delta H_{\mathrm{in}} + \omega\alpha_{\mathrm{eff}}/\gamma$, where $\Delta H_{\mathrm{in}}$ is the inhomogeneous contribution to the linewidth. $\alpha_{\mathrm{eff}}$ (blue diamonds) is fitted well by equation \ref{eq:alpha} (dashed line).
(b) Cobalt thickness dependence of the spin-pumping charge current is plotted (red circles). The relative size of the spin-current (solid blue line), which is calculated using the fits to the measured values of $M_{\mathrm{eff}}$ and $\alpha_{\mathrm{eff}}$, decreases in the thinner layers. In contrast, the charge current increases in the thinner layers. 
(c) The relative size of $\theta_{\mathrm{ISHE}}$ (red circles) is enhanced in the 1nm Co layer. The error bars show the standard error from fitting the $\sin\theta$ parameter to the angular-dependent symmetric voltage data. The small variance between the data points of the same thickness could also be from a small difference in the size of the microwave field in each device. The inhomogeneous part of the linewidth (blue diamonds) also shows an increase in thinner Co layers.}
 \end{figure}

We measured both $\alpha_{\mathrm{eff}}$ and $M_{\mathrm{eff}}$ with FMR. Values are shown in figure \ref{fig:current}a and are fitted well by equations \ref{eq:alpha} and \ref{eq:Meff} when $g_{\mathrm{eff}}^{\uparrow\downarrow}$ is constant for all the cobalt thicknesses, showing that there is no enhancement in the size of $j_{\mathrm{s}}^{0}$ with $d_{\mathrm{Co}}$.

The symmetric $\sin\theta$ voltage with the ISHE symmetry was converted to a DC current by dividing, for each device, by the individual resistance measured. Figure \ref{fig:current}b shows both the charge current for the different layers and the relative size of the spin current calculated from equation \ref{eq:js}.

The charge current generated in the device has a minimum at around 1.75 nm, whereas the spin-current decreases to zero as the ferromagnetic layer is reduced. The reproducibility of the results for each repeated measurement demonstrates that the increase in current in the thinnest layers cannot be attributed to variation in $h_{\mathrm{z}}$ between devices. The conversion of the interfacial spin-current to the charge current must therefore be dependent on the cobalt thickness. Our result is surprising as previous studies of thicker Py/Pt bilayers have shown remarkable agreement with the theoretical model\cite{PhysRevB.83.144402,PhysRevB.85.144408}. However, the minimum thickness of the ferromagnetic layer measured in those studies was 5 nm, significantly thicker than the range we have measured.

As $d_{\mathrm{Pt}}$ is the same for each device, the enhancement in the charge current observed must originate from a difference in the ISHE at the interface, and not the bulk ISHE in the Pt layer. The relative size of $\theta_{\mathrm{ISHE}}$, calculated from equation \ref{eq:V_ISHE} is plotted in figure \ref{fig:current}b and shows an enhancement of 2.4 times between the 2 nm and 1 nm Co layer. XRR measurements on a thicker reference bilayer show the substrate has a surface roughness of 0.42$\pm$0.07 nm. We expect thinner films grown on the substrate to retain more of this roughness, and consequently to have a larger Co/Pt interface region. This agrees with the inhomogeneous part of the linewidth shown in figure \ref{fig:current}c which indicates that the Co layer becomes increasingly less uniform in the thinnest films.

Recent studies have shown that surface and bulk impurities can greatly enhance the extrinsic SHE\cite{PhysRevLett.105.216401,PhysRevLett.109.156602} due to skew scattering. We expect there could be an enhancement to the ISHE in our samples at, or near, the interface caused by Co impurities in the Pt layer and vice versa. A rougher interface would lead to both a larger surface area with a greater number of surface impurities, and also a wider interface region containing impurities. 

Our experimental observation of the increase in the ISHE in ultra-thin layers motivates development of a theory modelling impurity scattering at rough interfaces. The enhancement observed allows more efficient conversion of spin to charge current in ultra-thin layers.

The authors would like to acknowledge useful discussions with Joerg Wunderlich, Gerrit Bauer and Jan Zemen. A.J.F acknowledges support from the Hitachi research fellowship and a Royal society research grant (RG110616). This work was partially funded by EPSRC grant EP/H003487/1.

\end{document}